# Specific Heat Discontinuity $\Delta C$ vs $T_c$ in Annealed $Ba(Fe_{1-x}Co_x)_2As_2$

J. S. Kim, B. D. Faeth, and G. R. Stewart

Department of Physics, University of Florida, Gainesville, FL 32611-8440

**Abstract:** The low temperature specific heat of annealed single crystal samples of $Ba(Fe_{1-x}Co_x)_2As_2$ with compositions spanning the entire superconducting phase diagram was measured. Effort was made to discover the best annealing schedule to maximize $T_c$ and minimize transition width in these samples. Values of $\Delta C/T_c$ normalized to 100% superconducting volume fractions varied proportionally to $T_c^{\alpha}$. Within a rather narrow error bar of $\pm 0.15$, the exponent $\alpha$ was the same (approximately 2) over a range of compositions ($0.055 \le x \le 0.15$) around the optimal concentration, x=0.08, where $T_c$ is the maximum in the phase diagram. Thus, whether the superconductivity was coexistent with magnetism (underdoped) or not (overdoped) did not affect the non-BCS variation ($\alpha$ nearly 2 instead of $\approx 0.8$-0.9 for BCS superconductors) of $\Delta C/T_c$ with $T_c$. The annealed samples in the present work, with increased $\Delta C/T_c$ and $T_c$ values compared to previous results for the $Ba(Fe_{1-x}Co_x)_2As_2$ alloy system, suggest that the *intrinsic* value of the exponent $\alpha$ for the iron superconductors, when the samples are annealed, with fewer defects, may indeed be even further from the BCS value of 0.8-0.9 than previously thought.

## I. Introduction

Bud'ko, Ni and Canfield[1] (2009) (hereafter 'BNC') discovered one of only a few known 'global' correlations in the new family of iron superconductors, joining bond angle and pnictogen height which are metrics[2] for predicting $T_c$ based on materials properties in this new class of materials. They found for 14 samples of various doped $BaFe_2As_2$ superconductors (including Co, Ni, Pd and Rh on the Fe site and K on the Ba site) that the discontinuity in the specific heat, $\Delta C$, at the superconducting transition temperature varies with $T_c$ as $\Delta C/T_c \propto T_c^\alpha$, $\alpha \approx 2$. Succeeding measurements on both additional iron pnictide and the iron chalcogenide (FePn/Ch) superconductors have tended to verify the BNC correlation, with $\alpha \approx 1.9$ (for a review see ref. 2.) This correlation between $\Delta C$ and $T_c$ in the FePn/Ch caused Kim et al.[3] to check if a similar comparison could be made for other families of superconductors, with the result that $\Delta C/T_c \propto T_c^\alpha$, $\alpha = 0.8 \pm 0.1$, for a wide collection of BCS, electron-phonon coupled superconductors as well as for some well known unconventional heavy Fermion superconductors. Even a subset of the high $T_c$ cuprate superconductors, which are also non-BCS, appears[2] to follow the slower variation of $\Delta C$ with $T_c$ like the BCS and heavy Fermion superconductors, $\Delta C/T_c \propto T_c^{0.8 \pm 0.1}$, leaving the FePn/Ch as evidently belonging to a different class of behavior.

Clearly, such a correlation may offer an insight into the fundamental pairing mechanism of the FePh/Ch superconductors. Various theoretical works[4-7] have addressed this. Kogan[5] derives $\Delta C/T_c \propto T_c^2$ for the case of strong pair breaking in a weak coupled BCS model, where the measured $T_c \ll T_{c0}$, the critical temperature of clean material. Whether the iron superconductors as a class are in the strong pair breaking limit is not clear. For example, P-doped

BaFe$_2$(As$_{0.67}$P$_{0.33}$)$_2$, T$_c$=30 K, which obeys[3] the BNC $\Delta C/T_c \propto T_c^2$ relationship, appears to be more in the clean limit based on, e. g., the ability to perform deHaas van Alphen measurements.[8] Zaanen[6] discusses the idea that the superconductivity could be forming from a non-Fermi liquid quantum critical metal. Evidence for quantum critical behavior has certainly been seen[2] in several iron superconducting systems. Vavilov and Chubukov[7] in their disorder model predict that away from the strong doping regime, where Kogan has shown that $\Delta C/T_c \propto T_c^2$, the dependence of $\Delta C/T_c$ on T$_c$ is more complex and differs from T$_c^2$. In particular, they predict that $\Delta C/T_c$ decreases *faster* with decreasing T$_c$ in the underdoped regime (where magnetism coexists with superconductivity) than in the overdoped regime. This is an interesting, and relatively easy to check, prediction.

The purpose of the present work is to present a consistent and more complete set of measurements on $\Delta C/T_c$ vs T$_c$ as a function of composition (nine compositions) in Ba(Fe$_{1-x}$Co$_x$)$_2$As$_2$ across the entire superconducting phase diagram. Ba(Fe$_{1-x}$Co$_x$)$_2$As$_2$ offers the advantage that samples made from self-flux, upon annealing, offer reproducible properties if made in a consistent fashion. In contrast, K-doped BaFe$_2$As$_2$ samples are difficult to prepare reproducibly, with the K-composition difficult to control.[2] Co-doped BaFe$_2$As$_2$ offers the advantage over Ni-doping of having a twice smaller rate of variation of T$_c$ with doping concentration, allowing for better control of T$_c$ with doping.

We will also compare our data on Ba(Fe$_{1-x}$Co$_x$)$_2$As$_2$ with the rather disjoint set of $\Delta C/T_c$ data already published (ref. 1, x=0.038, 0.047, 0.058, 0.078, 0.10, 0.114; ref. 9, x=0.075; ref. 10, x=0.04, 0.05, 0.055, 0.0575, 0.075, 0.09, 0.112, 0.12; ref. 11, x=0.045*, 0.08*, 0.103, and 0.105*, with '*' denoting samples annealed at 800 °C for 2 weeks). One[11] of these previous works began the study of property behavior as a function of annealing which we have built upon in the present work.

One of the difficulties in making a clear statement in this discussion is the nature of the $\Delta C$ data which are being correlated. As pointed out in, e. g., refs. 2 and 11, there are several difficulties in determining $\Delta C$ accurately. Perhaps less importantly, due to quality issues the samples to date in the FePn/Ch superconductors have sometimes quite broadened transitions, $\Delta T_c > 1$ K. As discussed in ref. 2, as long as the idealized superconducting transition (see below for graphical examples) is constructed to match entropies at $T_c$, the width of the transition does not seriously impede accurate (to ≈ 5%) determination of $\Delta C/T_c$.

Secondly, compositions of $Ba(Fe_{1-x}Co_x)_2As_2$ away from optimally doped, x=0.08, appear to have some significant fraction (sometimes exceeding 50%) of normal (or at least gapless) material below $T_c$ (evidenced by a finite $C/T$ rather than $C/T \approx 0$ in the superconducting state as T→0). This was first clearly pointed out in $Ba(Fe_{1-x}Co_x)_2As_2$ by Hardy et al.[10] (Analyzing the data of BNC[1] also shows the existence of this normal fraction, both for Co- and Ni-doped material.) Since $\Delta C$ at $T_c$ in a sample is from the gapped superconducting fraction, any precise discussion of $\Delta C/T_c$ vs $T_c$ as a function of composition <u>must</u> discuss the $\Delta C/T_c$ value normalized to a 100% gapped superconducting sample, see refs. 2 and 11 for discussions. This can be done if the normal state $\gamma_n$ (extrapolated from above $T_c$) and the residual $\gamma_r \propto$ non-superconducting/gapless fraction (extrapolated from low temperature data in the superconducting state as T→0) are known, with the normalized $\Delta C/T_c = \Delta C/T_c^{measured} * (\gamma_n/(\gamma_n - \gamma_r))$. This has been largely ignored in the more qualitative-in-nature previous discussions, e. g. in BNC[1] and Kim et al.[3].

The source of this residual $\gamma_r$ in $Ba(Fe_{1-x}Co_x)_2As_2$, and in the iron based superconductors in general, remains a matter of discussion. Thermal conductivity[12], $\kappa$, on $Ba(Fe_{1-x}Co_x)_2As_2$ shows a lack of a residual linear term $\kappa/T$ as $T\rightarrow 0$ in the superconducting state, implying the lack of connected normal metallic regions, whether from second phase or inhomogeneity in the majority phase. This leaves some sort of impurity-caused states in the gap or nanoscale inhomogeneity (as offered as an explanation for similar behavior in the high $T_c$ cuprates) as possible explanations. Whether such causes are *intrinsic* to $Ba(Fe_{1-x}Co_x)_2As_2$ (i. e. connected to the fundamental superconducting mechanism) or *extrinsic* (which would therefore become less important with an improved sample quality perhaps through improved annealing) is at present undecided. It is worth noting that a recent review[13] on the high $T_c$ cuprates concluded that the oft observed nanoscale inhomogeneities are "most likely due to inhomogeneous oxygen distribution" and that homogeneous samples "could be obtained by adequate annealing." Thus, at least in the cuprates, the review concludes[13] that the nanoscale inhomogeneities "are not an essential part" of the high temperature superconductivity mechanism.

As already mentioned, the residual $\gamma_r$ values grow (as was shown clearly in ref. 10, although as will be seen in the present work – see also ref. 11 - this effect is somewhat reduced in annealed samples) as Co composition, x, is varied from the optimal x=0.08 in $Ba(Fe_{1-x}Co_x)_2As_2$. Because of this, unrenormalized $\Delta C/T_c$ values tend to appear to fall *more rapidly* as x deviates further from 0.08 due to the decreasing amount of gapped superconducting fraction underlying the specific heat discontinuity. Thus, the slope previously found[1,3] for a wide variety of iron superconductor systems in $\log\Delta C/T_c$ vs $T_c$ plots of $\alpha\approx$1.9-2, where those plots were for as-measured, unrenormalized $\Delta C/T_c$ data, needs to be reevaluated in light of the respective $\gamma_r$ behavior on a system by system basis.

For the present work's annealed $Ba(Fe_{1-x}Co_x)_2As_2$ samples (where the finite $\gamma_r$ has a decreased effect since it is reduced due to the annealing), ignoring $\gamma_r$ and just plotting unrenormalized $\Delta C/T_c$ vs $T_c$ causes of the order of 15% increase in the exponent $\alpha$ for samples around the optimal, x=0.08 concentration with $T_c > 0.4\ T_c^{opt}$. We will discuss both raw, and normalized to 100% gapped superconducting fraction, $\Delta C/T_c$ plots for comparison.

It is worthwhile to discuss the error in the normalized $\Delta C/T_c$ values. The breadth of the transitions gives, with the idealized transition method (see ref. 2 for a discussion), an error $\approx$ ±5% as already mentioned. Secondly, since $T_c$ values for these $Ba(Fe_{1-x}Co_x)_2As_2$ can be over 20 K, extrapolating the normal state data to T=0 to obtain the normal state $\gamma_n$ (needed for calculating $\Delta C/T_c^{normalized} = \Delta C/T_c^{measured} * \gamma_n/(\gamma_n - \gamma_r))$ can be somewhat difficult. Hardy et al.[10] and Gofryk et al.[11] have done a careful job of subtracting the phonon contribution to the specific heat and obtaining rather accurate values for $\gamma_n$ for their samples. In addition, the specific heats of two of the samples in the present work (x=0.13 and 0.15) were measured to sufficiently high magnetic fields in another work[14] to determine $\gamma_n$ by suppressing the superconducting state with field. As will be seen from the listing of $\gamma_n$ values in Table 1 below, the variation of $\gamma_n$ according to all four studies (refs. 10-11, 14 and the present work) is rather slow as a function of composition in $Ba(Fe_{1-x}Co_x)_2As_2$. In particular, around the optimal composition (0.055 ≤ x ≤ 0.15) where $T_c$>10 K, $\gamma_n$ varies slowly vs x from approximately 17 $mJ/molK^2$ around x=0.05 and 0.15 up to approximately 21 $mJ/molK^2$ at the optimal composition around x=0.08. Thus, in calculating $\Delta C/T_c^{normalized} = \Delta C/T_c^{measured} * \gamma_n/(\gamma_n - \gamma_r)$, $\gamma_n$ for samples with $T_c$>10 K is known to approximately ±1 $mJ/molK^2$. As long as $\gamma_r$ is not too large a fraction of $\gamma_n$, this allows the total error bar for $\Delta C/T_c^{normalized}$ to not exceed $\approx$ ±10 %.

As made clear in ref. 3 in general, and by this discussion for $Ba(Fe_{1-x}Co_x)_2As_2$ in particular, the normal state specific heat $\gamma_n$ is *not* simply a constant independent of $T_c$ in these systems, or in any superconductor. However, the original BNC comparison[1] focused on $\Delta C/T_c$ vs $T_c$, not $\Delta C/\gamma T_c$ vs $T_c$, and this will be followed in the present work. Due to the decrease of $\gamma_n$ away from the optimal concentration in $Ba(Fe_{1-x}Co_x)_2As_2$ as just discussed, a plot of $\Delta C/\gamma T_c$ vs $T_c^{\alpha}$ for these materials would have a slightly smaller exponent $\alpha$. As discussed in Hardy et al.[10] (see also Table 1), the slight fall off of $\gamma_n$ away from $x_{opt}$ in $Ba(Fe_{1-x}Co_x)_2As_2$ is approximately symmetric about $x_{opt}$. Thus a comparison plot of $\Delta C/\gamma T_c$ vs $T_c^{\alpha}$ for underdoped vs overdoped samples would not reach a different conclusion about the relative rate of fall off of this ratio compared to a discussion of $\Delta C/T_c$ vs $T_c^{\alpha}$ as in the present work and in BNC.

Gofryk et al.[11] were the first to show the effect of annealing on the specific heat and $\Delta C$. They found (see Table 1) in their optimally doped $Ba(Fe_{0.92}Co_{0.08})_2As_2$ an increase in $\Delta C/T_c$ of 40% and an increase in $T_c$ of 25% after annealing the sample 2 weeks at 800 °C. The present work has investigated the effect of various annealing programs to optimize $T_c$ and transition width in $BaFe_{2-x}Co_xAs_2$. The result has both further improved $\Delta C/T_c$ over the results of ref. 11 and insured a set of samples in which the variation of the specific heat discontinuity with doping is quite reproducible.

**II.    Experimental**

Samples of $Ba(Fe_{2-x}Co_x)_2As_2$ (x=0.04, 0.055, 0.07, 0.08, 0.105, 0.13, 0.14, 0.15 and 0.19) were prepared using growth from FeAs self-flux. These compositions were chosen to give samples of comparable $T_c$ values on both sides (under- and overdoped) of $x_{opt}$. The samples were grown in outgassed $Al_2O_3$ crucibles welded into Nb containers. The Nb containers were heated to 1200 °C in flowing argon, held there for 4 hours, and then cooled at 3 °C/hour down to 900 °C, followed by cooling to room temperature at 75 °C/hour. Plate-like single crystals were easily mechanically separable from the flux. These crystals were then annealed for 1, 2, and 4 weeks at 600, 700, and 800 °C. 600 °C was not found to markedly change either $T_c^{onset}$ or the transition width, $\Delta T_c$. As an example of the annealing results for all the compositions, Fig. 1a shows the DC zero field cooled (H=10 G) results for optimally doped (i. e. the composition where $T_c$ is the maximum vs composition), x=0.08. Annealing at 700 °C produced an approximately 1 K increase in $T_c^{onset}$, as did annealing at 800 °C. However, the sharpness of the transition was optimized for 1 week annealing at 700 °C. Thus, the single annealing regimen used by Gofryk et al.[11] (2 weeks at 800 °C) was not far from the more optimal annealing schedule determined in the present work.

The resultant $T_c$ vs composition values for the annealed samples are shown in Fig. 1b. As mentioned, the compositions in the present work were chosen to give points on either side of the optimal doping concentration with comparable $T_c$'s, in order to carefully trace the dropoff in $\Delta C/T_c$ as x varies in the vicinity of $x_{optimal}$ on both the under- and overdoped sides. As will be discussed below when comparisons with results from other laboratories are made, it is important to note that the stated composition is not a good basis for exact inter-laboratory comparison. This is particularly true at and near the optimal composition where, as shown in Fig. 1b, $T_c$ is changing quite rapidly with small changes in the Co doping. Since the composition is an intrinsic parameter, and the parameters of interest are $\Delta C/T_c$ and $T_c$, this apparent variation in the actual composition between different laboratories is not a hindrance in determining if $\Delta C/T_c$ falls more rapidly with $T_c$ on the underdoped side of the phase diagram or not.

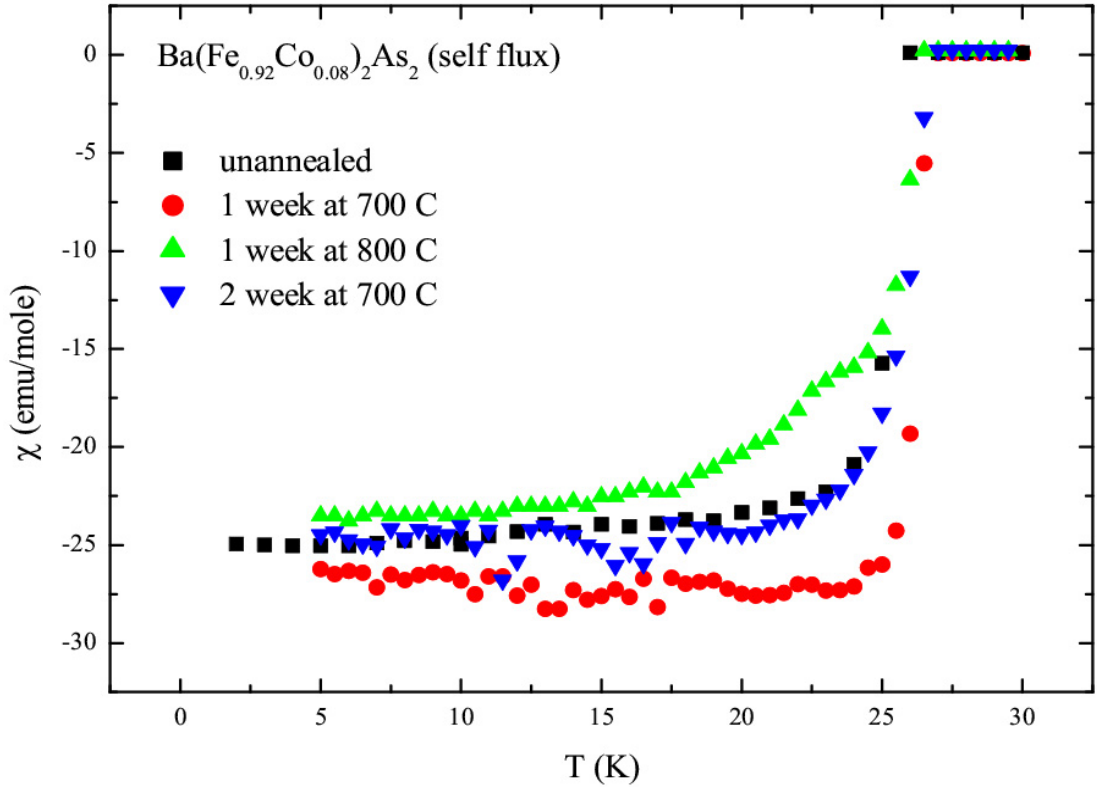

Fig. 1a. (color online) DC magnetic susceptibility of Ba(Fe$_{0.92}$Co$_{0.08}$)$_2$As$_2$ (so called "optimal" doping) for various annealing conditions. Data not shown for 4 week annealing at both 700 and 800 °C show broader $\Delta T_c$ and less shielding than for the two week annealing. Within the uncertainty of the demagnetization factor of the ~1.6 mg crystals measured here, the data for the samples shown here exhibit full diamagnetic shielding.

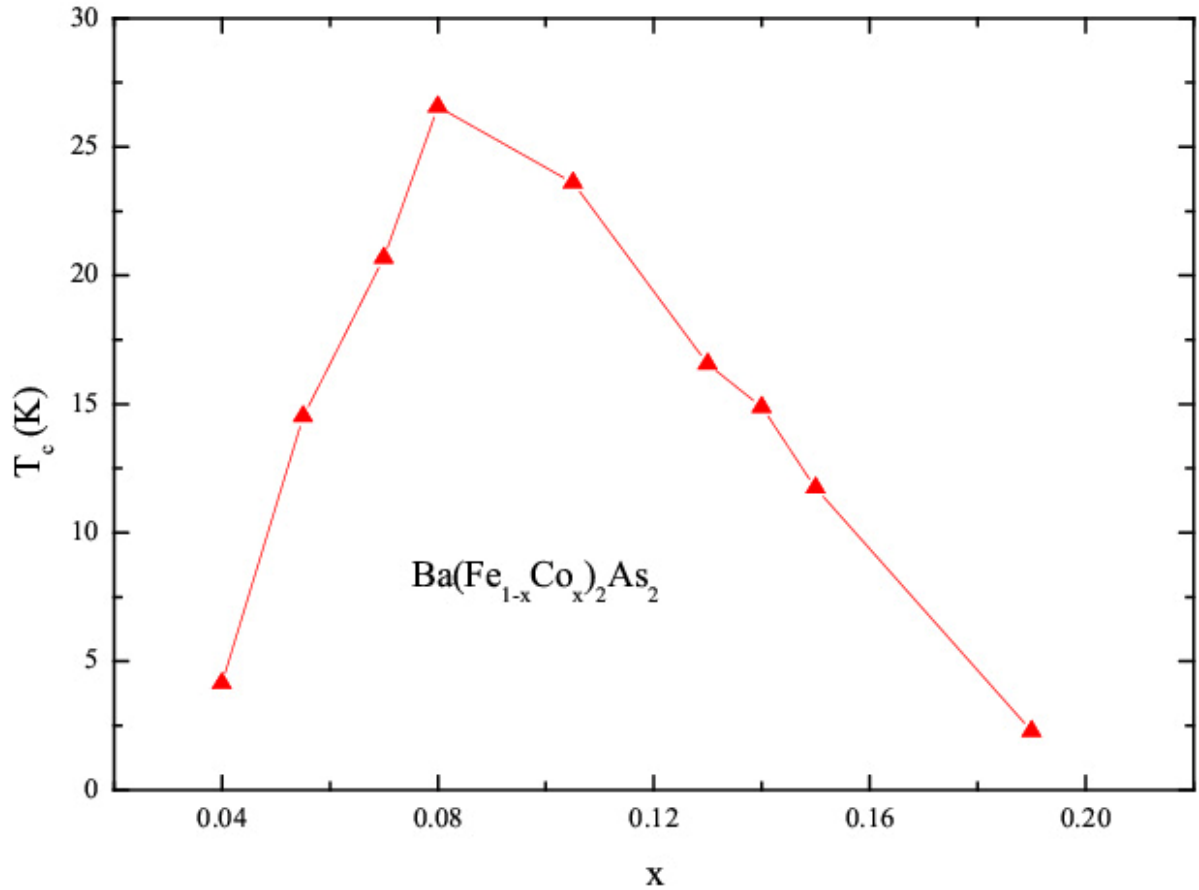

Fig. 1b. (color online) $T_c^{midpoint}$ determined from specific heat as a function of composition in Ba(Fe$_{1-x}$Co$_x$)$_2$As$_2$ for the samples annealed at 700 $^oC$ for one week. Note that the superconducting 'dome' in this phase diagram is not as rounded as typically (see, e. g., ref. 2) found for unannealed samples. These $T_c$ vs x data imply the possibility of even higher $T_c$ slightly above the nominal 8% concentration.

### III. Results and Discussion

The specific heats of annealed Ba(Fe$_{1-x}$Co$_x$)$_2$As$_2$, x= 0.04, 0.055, 0.07, 0.08, 0.105, 0.13, 0.14, 0.15, and 0.19, are shown in Fig. 2a. Fig. 2b gives an expanded view of annealed optimally doped Ba(Fe$_{0.92}$Co$_{0.08}$)$_2$As$_2$, while Fig. 2c shows an expanded view of annealed Ba(Fe$_{1-x}$Co$_x$)$_2$As$_2$, x=0.055 and 0.13 which have similar $T_c$ values. The numerical values of $T_c$, $\gamma_n$, $\gamma_r$, $\Delta C/T_c$, and the value of $\Delta C/T_c$ normalized for an idealized fully gapped superconducting sample where C/T would go to zero as T→0 (=$\Delta C/T_c^{measured} * \gamma_n/(\gamma_n-\gamma_r)$) for the samples shown in Fig 2a are listed in Table 1. As well, values from refs. 1 and 9-11 for the various compositions of Ba(Fe$_{1-x}$Co$_x$)$_2$As$_2$ measured in those works are also given where known. (Note that $\gamma_r$ values were not given for the samples measured in the original BNC work[1] but are reported here by scanning and analyzing the published low temperature C/T data.)

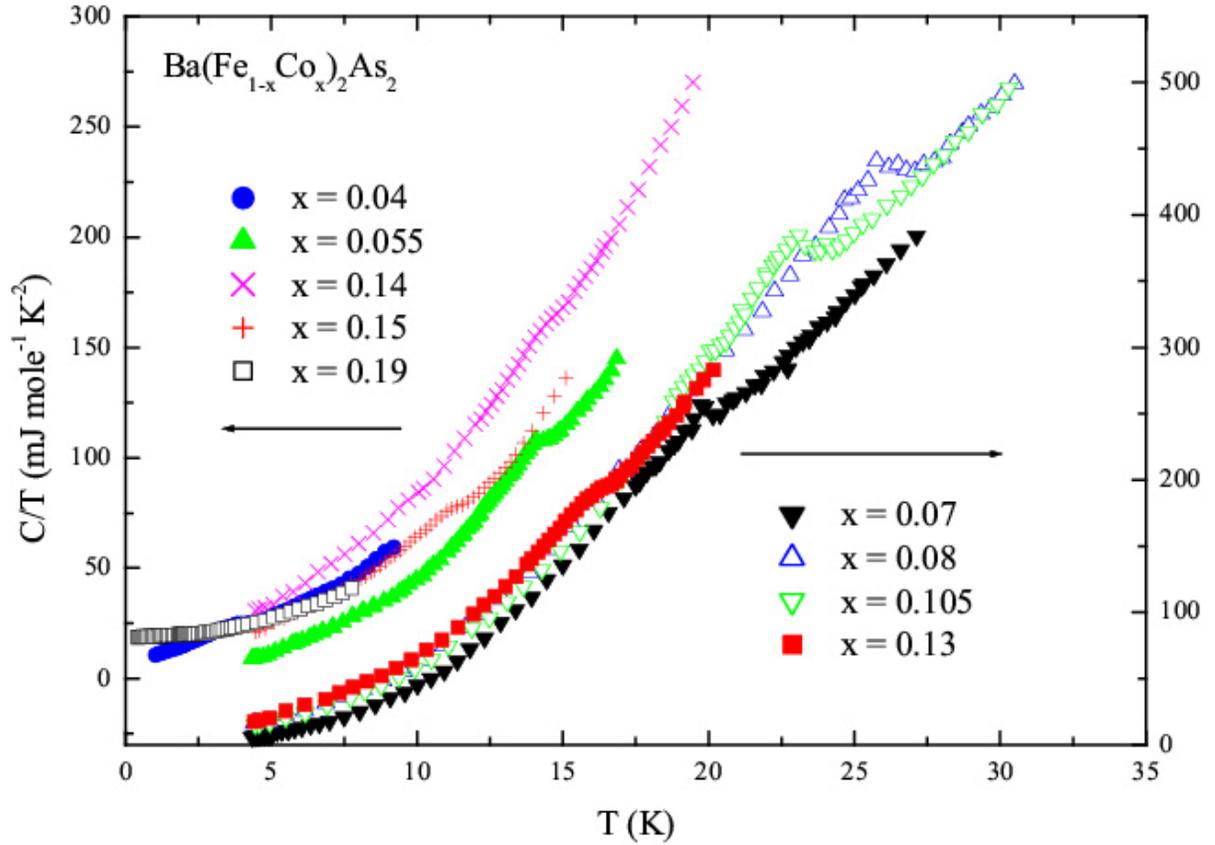

Fig. 2a (color online) Specific heat divided by temperature, C/T, vs temperature for annealed single crystals of Ba(Fe$_{1-x}$Co$_x$)$_2$As$_2$ for $0.04 \leq x \leq 0.19$. Note the different vertical axes for the two sets of curves. The x=0.19 sample (see Table 1) has a very small $\Delta C/T_c$ value and a rather large $\gamma_r$.

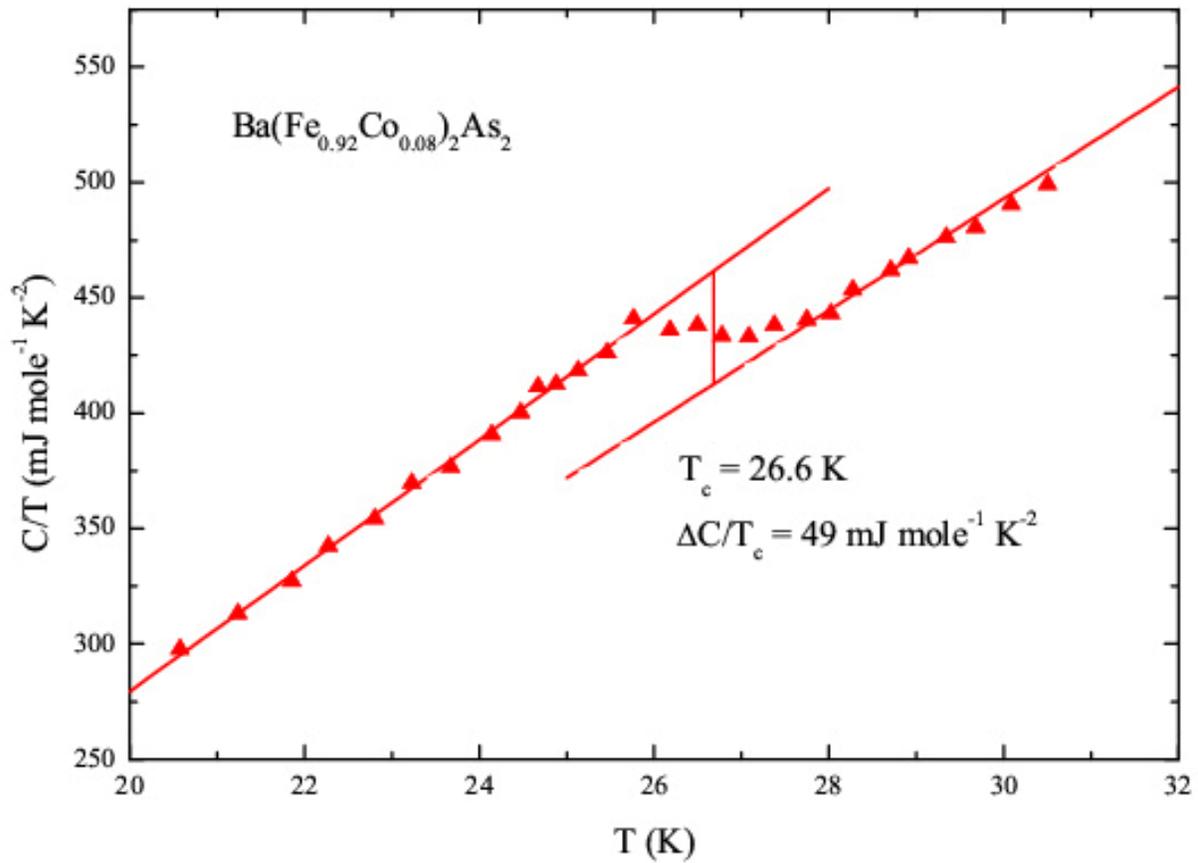

Fig. 2b (color online) Specific heat divided by temperature expanded around $T_c$ in annealed single crystal optimally doped $Ba(Fe_{1-x}Co_x)_2As_2$ to show detail in the specific heat discontinuity $\Delta C$. The transition width, $\Delta T_c$, even for this annealed sample is about 1.5 K, or 5% of $T_c$. As discussed in ref. 2, an idealized transition is formed to match the entropies at a perfectly sharp transition given by the vertical line. Note that this $T_c$ is a record high value for $Ba(Fe_{1-x}Co_x)_2As_2$.

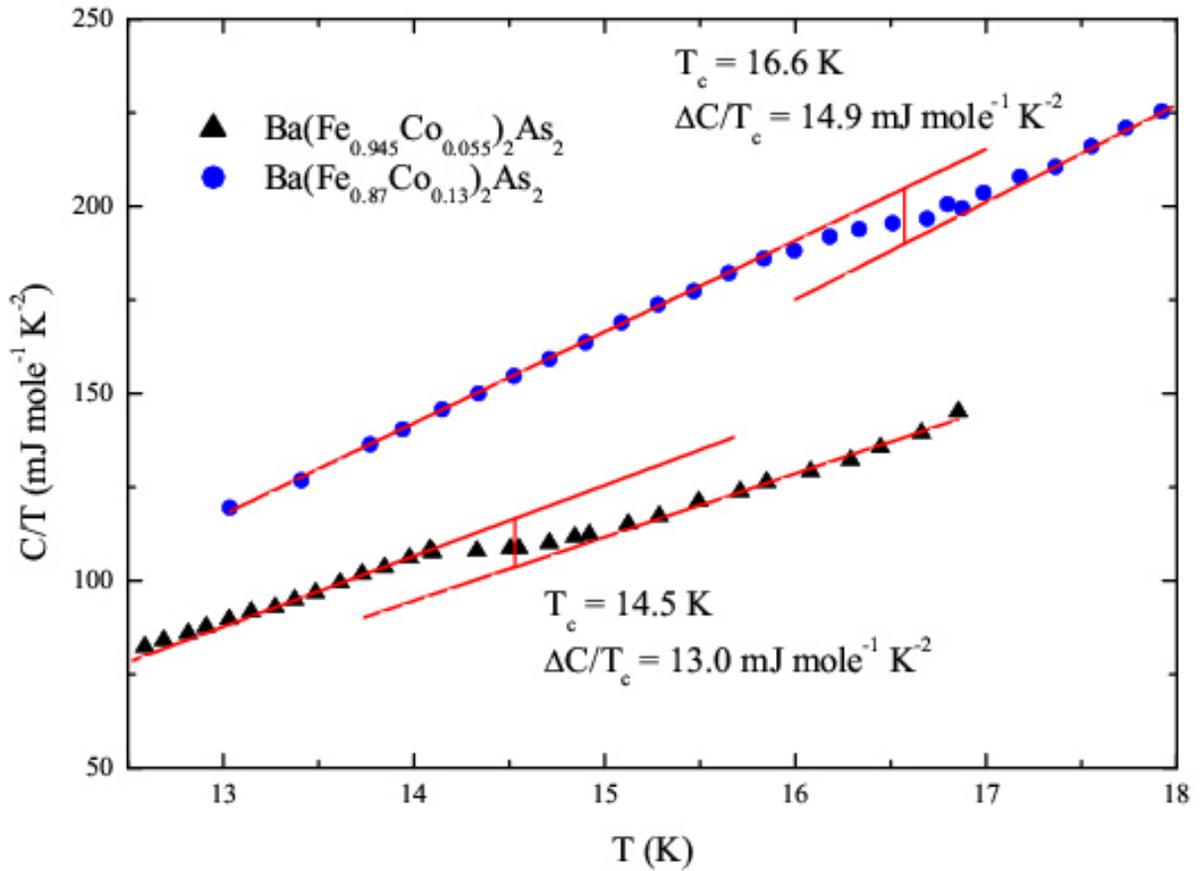

Fig. 2c (color online) Idealized transition ΔC constructed for x=0.055 and 0.13 in annealed single crystals of Ba(Fe$_{1-x}$Co$_x$)$_2$As$_2$. ΔT$_c$ values are less than 1 K. There are further normal state data above 18 K up to 20 K (not shown here, see Fig. 2a) for the x=0.13 sample which were included in the fit shown.

Before the results of the present work are discussed, we present in Fig. 3a a plot of raw, unrenormalized ΔC/T$_c$ vs T$_c$ values from refs. 1, 10, and 11. Filled symbols are for compositions above the optimal, maximum T$_c$ composition, i. e. for the overdoped samples, while 'X' denotes the optimal doping point and open symbols are for the underdoped side of the phase diagram. The average slope of logΔC/T$_c$ vs logT$_c$ for all three data sets is 1.56 (1.60 if the three annealed points from Gofryk et al. are omitted), which is smaller than the number of approximately 2 quoted by BNC[1] in their original paper (which included ΔC/T$_c$ data for Ba(Fe$_{1-x}$M$_x$)$_2$As$_2$, M=Co,

Ni, Pd, and Rh, as well as for $Ba_{1-x}K_xFe_2As_2$) but similar to the value of $\approx 1.5$ quoted by Hardy et al.[10] for their $Ba(Fe_{1-x}Co_x)_2As_2$ $\Delta C/T_c$ data. As discussed above, this value of $\alpha \approx 1.6$ will be steeper than an equivalent plot using $\Delta C/T_c$ renormalized to 100% gapped superconducting fraction, shown below.

One general observation for Fig. 3a is that below $T_c$=10 K (i. e. for compositions further from $x_{opt}$), there is more scatter in the $\Delta C/T_c$ values vs $T_c$. Second, as an example of the importance of $\gamma_r$, consider the underdoped $\Delta C/T_c$ point at $T_c$=8.0 K from Gofryk et al. in Fig. 3a. Note that this point (see numerical values in Table 1) appears higher than the trend of the other data in that $T_c$ range (for instance the overdoped, filled symbol from Bud'ko et al. at $T_c$=8.5 K). However, this is a misleading comparison, because since this point is from an annealed sample, it has a lower residual $\gamma_r$ – from Table 1 $\gamma_r$ for this Gofryk et al. $T_c$=8.0 K sample is only 1.0 mJ/molK$^2$ vs $\gamma_r$=11.1 mJ/molK$^2$ for the $T_c$=8.5 K BNC overdoped sample. Thus, the unrenormalized, as-measured $\Delta C/T_c$ value for the annealed, lower $\gamma_r$ Gofryk et al. sample is from a larger percentage of superconducting sample and is thus larger than, e. g., the BNC point at $T_c$=8.5 K where the measured $\Delta C/T_c$ value, with a larger $\gamma_r$ value, is from only about half the sample and is therefore smaller.

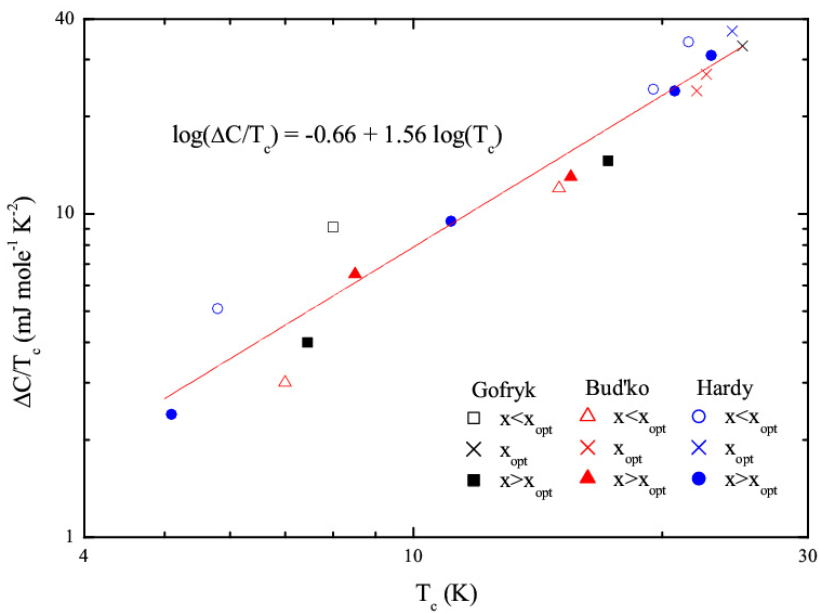

Fig. 3a (color online) Unrenormalized values of $\log(\Delta C/T_c)$ vs $\log(T_c)$ for single crystals of $Ba(Fe_{1-x}Co_x)_2As_2$ from refs. 1 (Bud'ko, Ni, and Canfield), 10 (Hardy et al.) and 11 (Gofryk et al.) For the data from BNC, ref. 1, two values of x (0.058 and 0.078) have almost identical $T_c$'s (22.6 and 22 K respectively), so that both $\Delta C/T_c$ values are denoted by 'X' in the figure. Note that in these unrenormalized $\Delta C/T_c$ data, with samples from different laboratories and one set (Gofryk et al.) being annealed while the other two sets are not, there does not appear to be a significant difference for $T_c>10$ K (where $\gamma_r$ is less than $0.5\gamma_n$ and there is less scatter) in the slope of $\log\Delta C/T_c$ with $\log T_c$ for underdoped (open symbols) vs overdoped (closed symbols) samples.

However, as discussed, it is clear that in order to accurately investigate whether $\Delta C/T_c$ falls faster for underdoped than for overdoped iron superconductor samples as composition varies from $x_{opt}$ in $Ba(Fe_{1-x}Co_x)_2As_2$, it is necessary to consider $\Delta C/T_c$ values adjusted for the fraction of the sample that is a gapped superconductor. This is not just because there are widely varying amounts of superconducting fraction at different compositions, even concentrating on $T_c>10$ K (e. g. from over 90% for x=0.07 to only about 50% for x=0.13 in the present work), which prevents valid intercomparison of the variation of $\Delta C/T_c$ values at different $T_c$ (and composition) values. In addition, the variation in $\gamma_r$ between samples of the *same* composition can be significant between, or even within, laboratories. For example, in optimally doped $Ba(Fe_{1-x}Co_x)_2As_2$, where $\gamma_r$ is small, $\gamma_r$ varies (see Table 1) between 1.4 and 6.6 mJ/molK$^2$ for the annealed samples from ref. 11 and the present work. Within different batches of optimally doped $Ba(Fe_{1-x}Co_x)_2As_2$ in one laboratory[11], $\gamma_r$ varies between 1.3 and 0.25 mJ/molK$^2$.

Thus, in Fig. 3b values (shown in Table 1) for the normalized $\Delta C/T_c$ in unannealed $Ba(Fe_{1-x}Co_x)_2As_2$ calculated from the data of Bud'ko, Ni, and Canfield[1] and Hardy et al.[10] and in annealed $Ba(Fe_{1-x}Co_x)_2As_2$ calculated from the data of Gofryk et al.[11] and the present work are plotted. Values of $\Delta C/T_c$ for samples with $\gamma_r>0.5\ \gamma_n$ normalized to $\Delta C/T_c * \gamma_n/(\gamma_n-\gamma_r)$ are not shown in Fig. 3b (but see italicized values in Table 1) because the factor $\gamma_n/(\gamma_n-\gamma_r)$ needed for $\Delta C/T_c^{normalized}$ magnifies the rather small error discussed above of $\pm 1$ mJ/molK$^2$ in $\gamma_n$ the larger $\gamma_r$ is.

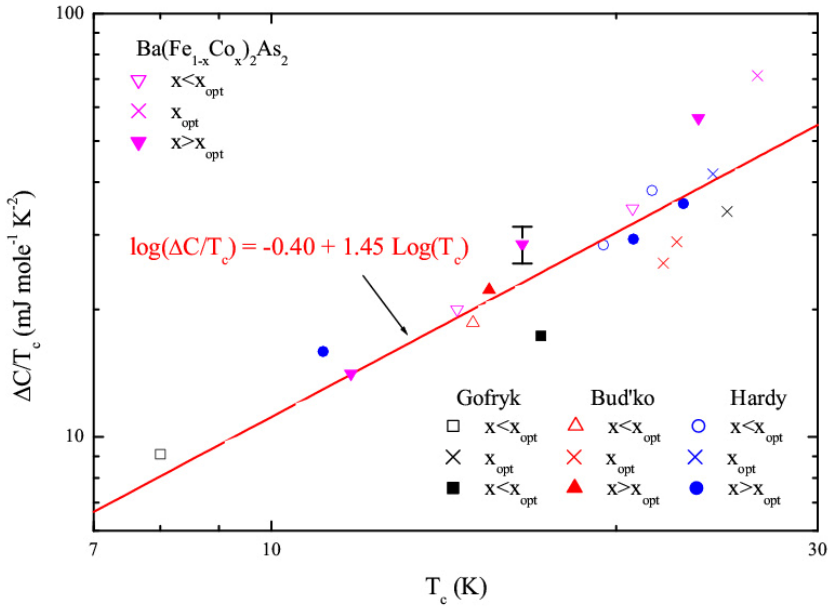

Fig. 3b (color online) Normalized values of $\log(\Delta C/T_c)$ vs $\log(T_c)$ for $Ba(Fe_{1-x}Co_x)_2As_2$ calculated from the data from refs. 1, 10-11 and the present work for compositions with $\gamma_r < 0.5\, \gamma_n$. The optimal $x \approx 0.08$ is denoted by 'X'. The normalized $\Delta C/T_c$ is given by $\Delta C^{measured}/T_c * (\gamma_n)/(\gamma_n - \gamma_r)$ and varies as $T_c^{1.45}$. The error bar shown is $\pm 10\,\%$, as discussed in the text.

In any case, a majority of these high $\gamma_r$ samples omitted from Fig. 3b have $T_c$ values less than 0.3 $T_c^{opt}$. This is presumably in the $T_c \ll T_{c0}$ limit (where $T_{c0}$ is for the ideal clean limit) where Kogan predicts $\Delta C/T_c \propto T_c^2$, and are not in the regime focused on by Vavilov and Chubukov where the relative rate of fall of $\Delta C/T_c$ with $T_c$ for underdoped and overdoped samples has been predicted to differ.

The 19 normalized $\Delta C/T_c$ values shown in Fig. 3b, even with the eight large $\gamma_r$ samples excluded, still show a large amount of scatter from the average fit line of $\Delta C/T_c \propto T_c^\alpha$, $\alpha \approx 1.45$. If the data from the annealed samples of ref. 11 and the present work (where annealing as will be discussed tends to increase the slope) are omitted, there is significantly less scatter as shown in Fig. 4a. A fit to these data from the works of BNC[1] and Hardy et al.[10] gives $\alpha \approx 1.13$. Although neither the $\Delta C/T_c$ data from BNC or Hardy et al. alone allow a conclusion on the slope $\alpha$ in under- vs overdoped $Ba(Fe_{1-x}Co_x)_2As_2$, together these unannealed, renormalized data in Fig. 4a show no obvious difference in the relative value of α. Another conclusion is that the slope of log$\Delta C/T_c$ vs log$T_c$ for $\Delta C$ values normalized to the fraction of the sample in Fig. 4a that is a gapped superconductor is indeed smaller than the exponent of 1.6 from the unrenormalized data of Fig. 3a as expected.

It is interesting to note that this value of α≈1.1 for renormalized $\Delta C/T_c$ for unannealed Ba(Fe$_{1-x}$Co$_x$)$_2$As$_2$ is not significantly different from the value obtained by Kim et al.[3] for BCS superconductors, where $\gamma_r$ for those data is negligible. However, a consideration of annealing

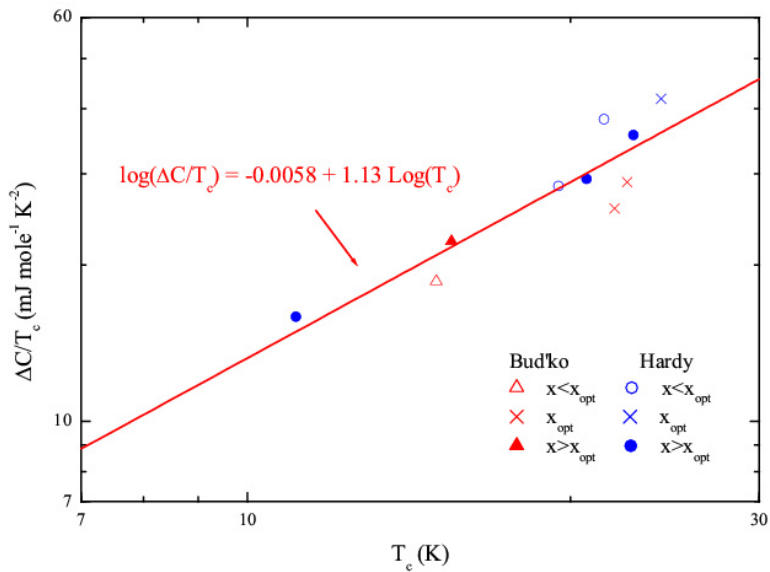

Fig. 4 a (color online) Normalized values of log($\Delta C/T_c$) vs log($T_c$) for Ba(Fe$_{1-x}$Co$_x$)$_2$As$_2$ calculated from the results of BNC[1] and Hardy et al.[10] for compositions near (down to $T_c$=11.1 K) to the optimal x=0.08 (which composition is denoted by 'X'). The normalized $\Delta C$ is given by $\Delta C^{measured} * (\gamma_n)/(\gamma_n - \gamma_r)$. The error bar shown is ±10 % as discussed in the text.

(which as discussed with Fig. 3b tends to increase the slope) still results in considering the iron superconductors as a separate class, as will now be discussed.

Having plotted only the unannealed renormalized $\Delta C/T_c$ from Fig. 3b in Fig. 4a, we now discuss the annealed data. Since the work of Gofryk, et al.[11], with only three $\Delta C/T_c$ values, has insufficient data to determine an α by itself, and since their annealed values (as discussed, see also Fig. 3b and Table 1) are significantly different from those of the present work due to the different sample preparation and annealing, we present here the self-consistent data from the present work. A careful consideration of samples made within one laboratory in a consistent fashion, as discussed in the Introduction, offers the best opportunity to decide the question raised by Vavilov and Chubukov.

Fig. 4b shows the normalized $\Delta C/T_c$ values vs $T_c$ for six annealed compositions with $\gamma_r < 0.5\,\gamma_n$ of $Ba(Fe_{1-x}Co_x)_2As_2$ from the present work. Although annealing has not made the superconducting phase transition ideally sharp ($\Delta T_c$ values, e. g. for x=0.08 in Fig. 2b, are still as wide as 1.5 K), the consistent process by which the samples were made and annealed provides reduced scattering in the results.

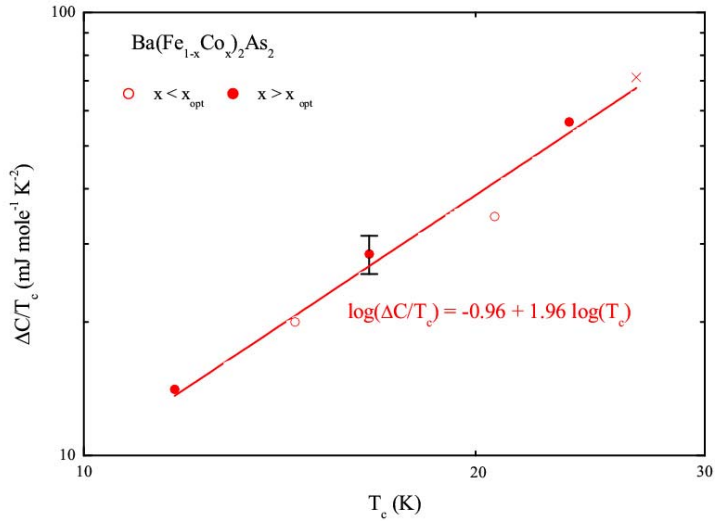

Fig. 4b (color online) Normalized values of $\log(\Delta C/T_c)$ vs $\log(T_c)$ for $Ba(Fe_{1-x}Co_x)_2As_2$ for annealed compositions near (down to $T_c$=11.7 K) to the optimal x=0.08 (which composition is denoted by 'X') from the present work. The normalized $\Delta C$ is given by $\Delta C^{measured} * (\gamma_n)/(\gamma_n - \gamma_r)$. The error bar shown is ±10% as discussed in the text. The sample (see Table 1) with x=0.14, $T_c$=14.5 K is omitted due to the large $\gamma_r$ of ≈90 % of $\gamma_n$. A plot of the 7 unrenormalized data points (not shown) $T_c$>10 K gives $\alpha$=2.30 and also shows no significant slope difference, under- vs overdoped..

Clearly, from Fig. 4b, the relative fall off of $\Delta C/T_c^{normalized}$ with $T_c$ decreasing away from $T_c^{opt}$ for under- and overdoped $Ba(Fe_{1-x}Co_x)_2As_2$ appears within a rather narrow amount of scatter to be identical. This conclusion, based on the data from the present work with less scatter, is in agreement with the unannealed results, Fig. 4a, of BNC[1] and Hardy et al.[10] Thus, the proposal of ref. 7, that the presence of the spin density wave magnetism on the underdoped side of the superconducting dome in iron based superconductors should cause $\Delta C/T_c$ to fall faster with $T_c$ decreasing away from $T_c^{optimal}$ than on the overdoped side, appears to not be borne out by the experimental results of the present work.

It is interesting to discuss the reason for the different slopes of $\log\Delta C/T_c^{normalized}$ with $\log T_c$ for the collection of four group's data in Fig. 3b with that shown in Fig. 4b ($\Delta C/T_c^{normalized} \propto T_c^\alpha$, $\alpha$=1.45 and 1.96 respectively.)  The data in Fig. 4b from the present work have a higher slope vs $T_c$ because – due to the optimized annealing carried out in the present work – firstly, the $\Delta C/T_c^{normalized}$ value at the optimal composition of 0.08 is significantly higher (factor of two) than that in the unannealed works of BNC[1] and Hardy et al.[10]. (The present work's $\Delta C/T_c$ values for optimal doping, both unadjusted and normalized to 100 % superconducting fraction, are also larger, >50 %, than those for the annealed at two weeks at 800 °C x=0.08 sample from Gofryk et al.)  As well, the present work's $\Delta C/T_c^{normalized}$ values (as shown in Figs. 3b and 4b and in Table 1) for the other compositions are higher than those of BNC and Hardy et al.  When plotted, these data result in a higher slope in the present work.

This of course represents a further challenge in trying to understand the intrinsic value of $\alpha$ in $\Delta C/T_c$ vs $T_c^\alpha$ for the broad range of iron superconductors considered in refs. 1 and 3 for unrenormalized values.  Not only should those values be adjusted for superconducting fraction ($\gamma_r$ is only about 10 % of $\gamma_n$ in, e. g., $BaFe_2(As_{0.67}P_{0.33})_2$[3], $Ba_{0.6}K_{0.4}Fe_2As_2$[15] and $FeSe_{0.88}$[16]) but the question raised by the present work is also – how optimized are the samples?  In the present work, we have shown that both unadjusted and normalized values of $\Delta C/T_c$ (and also $\alpha$) grow substantially with optimized annealing (in addition to the increase of $T_c^{opt}$ from 20-23 K in unannealed samples to 26.6 K for the annealed sample in the present work.)  This annealing increases the $\alpha$ in the renormalized $\Delta C/T_c$ vs $T_c^\alpha$ data from the present work to $\alpha$ almost equal to 2 from the lower value found in unannealed samples, 1.1.

Thus, for at least one example of the iron superconductors (Ba(Fe$_{1-x}$Co$_x$)$_2$As$_2$), annealing moves $\alpha$ further away from the behavior of $\Delta C/T_c \propto T_c^\alpha$, $\alpha \approx 0.8$-$0.9$, found in ref. 3 for a large number of BCS superconductors (where, since $\gamma_r$ values are typically $\approx 0$, no renormalization for a valid comparison with the present work is required).

**Conclusions**

A careful specific heat study of annealed Ba(Fe$_{1-x}$Co$_x$)$_2$As$_2$ samples over a wide range of composition was undertaken. This study reveals that, within a rather narrow error bar, $\Delta C/T_c \propto T_c^\alpha$ behaves the same on *both* the under- and overdoped sides of the optimal, maximum $T_c$ concentration for $T_c/T_c^{opt} > 0.4$. The resultant value of $\alpha$ is approximately 2 for $\Delta C/T_c$ normalized to 100 % gapped superconducting sample fraction and $\alpha \approx 2.3$ for unadjusted $\Delta C/T_c$. The present work argues that the proper way to consider $\Delta C/T_c$ vs $T_c$ is by taking into account the residual $\gamma_r$ value, and thus the gapped superconducting fraction of the sample that contributes to the specific heat discontinuity. However, both methods of considering $\Delta C/T_c$, and even a compilation of previously published data for unannealed samples, leads to the conclusion that the presence or absence of magnetism coexistent with the superconductivity does not have a measurable effect on the behavior of $\Delta C/T_c$ with $T_c$ in Ba(Fe$_{1-x}$Co$_x$)$_2$As$_2$. There is no reason[2,17-18] to presume that this compound is not representative of iron superconductors in general. Further, the exponent $\alpha$ in $\Delta C/T_c \propto T_c^\alpha$ in the optimally annealed (i. e. presumably more characteristic of *intrinsic* behavior) Ba(Fe$_{1-x}$Co$_x$)$_2$As$_2$ samples of the present work appears to be indeed different from that in phonon mediated (BCS) superconductors. As a metric for the quality of the samples, the present work has achieved a record $T_c$, 26.6 K, as measured by the midpoint of the specific heat transition, for single crystals of $Ba(Fe_{1-x}Co_x)_2As_2$. Work is underway using finer gradations in the Co-concentration above x=0.08 to search for the true maximum $T_c$ using the optimized annealing procedure developed in the present study.

**Acknowledgements:** The authors acknowledge helpful discussions with Kris Gofryk and Filip Ronning. Work at Florida performed under the auspices of the US Department of Energy, Basic Energy Sciences, contract. no. DE-FG02-86ER45268.

**Table 3 (color online):** Specific heat $\gamma_n$, $\gamma_r$, $\Delta C/T_c$ (measured and normalized for gapped superconducting fraction) and $T_c$ for unannealed and annealed*

$Ba(Fe_{1-x}Co_x)_2As_2$

| x= | $T_c$(K) | $\gamma_n/\gamma_r$ (mJ/moleK$^2$) | $\Delta C/T_c$ (mJ/moleK$^2$) | $\Delta C/T_c^{nor}$(mJ/moleK$^2$) | Ref. |
|---|---|---|---|---|---|
| 0.038 | 7 | -/10.3 | 3 | | BNC |
| 0.04 | 5.8//4.15* | 14.9/9.8//17*/9.2* | 5.1//4.3* | *14.9//9.4* | a//pw |
| 0.045 | 5.6,8.0* | 13.7,14*/10.2,1.0* | ~2-3/8.4* | 9.0* | b |
| 0.047 | 15 | ~16/5.7 | 12 | 18.6 | BNC |
| 0.05 | 19.5 | 17.2/2.5 | 24.3 | 28.4 | a |
| 0.055 | 21.5//14.5* | 19/2.1//19*/6.3* | 34/13* | 38.2//19.4* | a//pw |
| 0.0575 | 24.3//22.6 | 21.3/2.6//~21/1.4 | 36.7//27 | 41.8//28.9 | a//BNC |
| 0.07 | 20.68* | 21*/1.6* | 32.0* | 34.6* | pw |
| 0.075 | 22.9, 21.4 | 22.1/2.9, 23.8/5.77 | 30.9,29.2 | 35.6, 38.5 | a, a' |
| 0.078 | 22 | ~21/1.0 | 24.5 | 25.7 | BNC |
| 0.08 | 20,25*//26.57* | 18,22*/3.7,1.4* //21*/6.6* | 21.6,31.9*//48.9* | 27.2/34.1*//71.3* | b//pw |
| 0.09 | 20.7 | 20/3.6 | 24 | 29.3 | a |
| 0.10 | 15.5 | ~20/9.2 | 12 | 22.2 | BNC |
| 0.105 | 11,17.2*//23.6 | 23.2,20*/14.5,3.8* //21*/5.9* | 9.5/14* //40.7* | *17.7*/17.3*//56.6* | b//pw |
| 0.112 | 11.1 | 17/7.9 | 8.5 | 15.9 | a |
| 0.114 | 8.5 | -/11.1 | 6.5 | | BNC |
| 0.12 | 5.1 | 14.6/10.4 | 2.4 | *8.3* | a |
| 0.13 | 16.57* | 18*/8.6* | 14.9* | 28.5* | pw |
| 0.14 | 7.45//14.48 | NR /0.4*//?/18.3* | NR//10.2* | | b//pw |
| 0.15 | 11.74 | 18*/6.5* | 9.0* | 14.1* | pw |
| 0.155 | 0 | 16 | | | a |

| | | | | | | |
|---|---|---|---|---|---|---|
| 0.19 | 2.28 | 18.1*/18.7* | 0.6* | | | pw |

Ref. a: Hardy et al., ref. 10; Ref. a': Hardy et al., ref. 9; Ref. b: Gofryk et al., ref. 11 (note that the unannealed values from this work listed here are not plotted in Fig. 3a), 'NR' for x=0.14 means not reported; Ref. BNC: Bud'ko, Ni and Canfield, ref. 1; Ref. pw: values from the present work. Compositions for BNC[1] were analyzed using wavelength dispersive x-ray spectroscopy, for Hardy et al.[9-10] using energy dispersive x-ray spectroscopy, and for Gofryk et al.[11] using electron microprobe. Compositions for samples from the present work are nominal. As can be seen from the data, there is a significant variation in $T_c$ for samples of similar composition. Annealed values are with *. Values of normalized $\Delta C/T_c$ in italics have $\gamma_r > 0.5 \gamma_n$ and are not shown in the figures due to their larger error bar as discussed in the text. As may be seen for x=0.08, Gofryk et al. see a larger increase in $T_c$ with annealing than the 1 K increase seen in the present work and discussed in the text and in Fig. 1a. Thus, the unannealed x=0.08 samples of the present work, see Fig. 1a, started out with a higher $T_c$ than the unannealed x=0.08 samples in the work[11] of Gofryk et al., which reports a lower (by 1. 6 K) annealed $T_c$ for the optimal concentration than found in the present work.